\documentclass[10pt,twocolumn]{article}
\usepackage{isqcmc2025} 

\usepackage{graphicx,url}
\usepackage{braket}
\usepackage{hyperref}

\usepackage[latin1]{inputenc}

\usepackage{anonymize} 
\usepackage{listings}

\lstdefinelanguage{OPENQASM}{
morekeywords={q, c}, 
morekeywords=[2]{OPENQASM, include}, 
emph={h,cx,qreg,creg,->}, 
morekeywords=[3]{measure}, 
morekeywords=[4]{barrier}, 
sensitive=true,
morecomment=[l]{//}, 
morestring=[b]",
literate={->}{{\textbf{\color{codeemph2}{$\to$}}}}1 
}

\usepackage{orcidlink}

\title{Techniques for Quantum-Computing-Aided Algorithmic Composition:\\ 
\large Experiments in Rhythm, Timbre, Harmony, and Space}

\author{\anonymize{Christopher Dobrian\orcidlink{0009-0001-2757-8991}
}\inst{1} \and
    \anonymize{Omar Costa Hamido\orcidlink{0000-0001-5077-853X}
    }\inst{1}\inst{2}}

\address{\anonymize{University of California, Irvine CA 92697-2775, USA}
    \nextinstitute
    \anonymize{Centre for Interdisciplinary Studies (CEIS20), University of Coimbra, Portugal}
    \email{\anonymize{dobrian@uci.edu, ocostaha@uci.edu}}
}

\begin{document}

\maketitle

\begin{abstract}
Quantum computing can be employed in computer-aided music composition to control various attributes of the music at different structural levels. This article describes the application of quantum simulation to model compositional decision making, the simulation of quantum particle tracking to produce noise-based timbres, the use of basis state vector rotation to cause changing probabilistic behaviors in granular harmonic textures, and the exploitation of quantum measurement error to cause noisy perturbations of spatial soundpaths. We describe the concepts fundamental to these techniques, we provide algorithms and software enacting them, and we provide examples demonstrating their implementation in computer-generated music.
\end{abstract}

\section{Introduction}\label{sec:intro}
The process of music composition can be thought of as the process of making decisions. Lots and lots of decisions. At every moment, a composer chooses a sound---a musical note or a sonic event---from among the nearly infinite variety of all possible sounds. In order to manage this \textit{embarras de choix}, a composer necessarily self-imposes restrictions on the range of possibilities s/he will employ. A composer achieves these compositional constraints in any of a number of ways: using a formal rule-based system, or known theoretical guidelines, or implicit stylistic traits and clichés, or even deliberate arbitrariness or explicit randomization, to guide the dizzyingly large number of choices to be made in a composition.

The study of compositional decision making, and the design of algorithms that explain, elucidate, or even reinvent the compositional decision making process, comprise a field generally called \textit{algorithmic composition}. In the composer's thought process, the compositional ``algorithm'' may be well defined or it may be completely intuitive. For a listener hearing a new composition for the first time, however, the compositional process is largely mysterious, perhaps even irrelevant to the appreciation of the music. For the listener, every new note or sound is unknown until it occurs. Only once the next sound is heard can the listener attempt to cognize and comprehend that sound's relation to what has occurred before it, and perhaps develop some probabilistic expectation of what might come next. In short, every musical moment is unknown until it is discovered. But that process of discovery is different for the composer and listener. The composer or improviser discovers by making and enacting a decision, whereas the listener discovers by hearing the result of that decision. Because the composer has free will, \textit{anything} can theoretically come next in a piece of music. Every future musical moment is unknown, even if some things may be more probable than others.

This inherent unknowability of the next sonic instant can be likened to the state of \textit{superposition} in quantum theory. A sound in the present moment, and that sound's implications for the next moment, can be compared metaphorically to the superposition of a particle in quantum mechanics, or of a qubit in quantum computing. Starting from this simple metaphor of superposition representing the inherent unknowability of the next musical moment, and specifically representing the composer's or improviser's next decision, we have been using quantum circuit emulations to model compositional decision making. This article reports on some of those experiments, using qubits to make decisions at different levels of the algorithmic composition process. We will address instances of quantum circuits in the generation of rhythm, timbre, harmonic progression, and spatialization, especially in the context of highly contrapuntal and granular musical textures. Thinking of compositional choices in terms of quantum theory---quantum particle measurements or quantum computing techniques---has led us to some new musical ideas. Implementing compositional algorithms using quantum circuit emulations has produced some interesting experimental results, sounds which are somewhat different from those produced with conventional processes.

\section{Rhythm}\label{sec:rhythm}
At the most fundamental level, rhythm can be defined as the organization of sonic events in time. In order for us to perceive sounds as explicitly rhythmic, however, we must be able to discern patterns in that organization. Indeed, one might say that organization of things (in this case, sound objects) necessarily implies establishing patterns and/or some sort of immanent systemic logic in their placement. And in order for us to detect a pattern, there must exist some sort of repetition, either exact or similar, that enables comparison.

\begin{figure}[h]
\centering
\includegraphics[width=.15\textwidth]{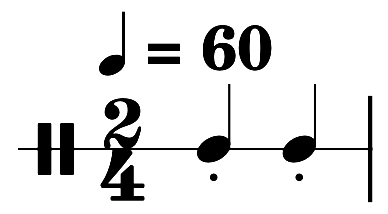}
\caption{Two timepoints}
\label{fig01}
\end{figure}

Central to the appreciation of rhythmic music is humans' innate mental ability to measure short periods of time\footnote{For the purpose of this discussion, referring to time periods commonly thought of as rhythmic units, we mean roughly $\frac{1}{8}$ of a second to 2 seconds.} with impressive accuracy, and then to extrapolate from those measurements to predict future events, and even to add or divide time periods mentally with good accuracy and precision. For example, imagine that you, the listener, hear the following, at almost any reasonable musical tempo \ref{fig01}. Two timepoints are established by the time interval between onsets of the two sounds, and almost anyone can predict the likely occurrence of the next sound on the downbeat of the next measure, and can even imagine what eighth notes or triplet eighth notes would sound like, based merely on the inter-onset interval (IOI) established by those two sounds.

The composer utilizes that mental time-measurement ability of the listener, in order to establish concepts such as beats (regularly spaced conceptual time intervals), beat divisions (rational fractions of those time intervals), and meter (groupings of beats), and to organize the occurrence of patterns in those features. Once a pattern has been established, by even as little as a single repetition of a single feature, expectations have been created for the listener, perceived likelihoods, however tentative. And once expectations have been created, the composer can decide to fulfill or disappoint those expectations for aesthetic purposes. In rhythm, this can take the form of syncopation (intentional displacement of a timepoint from its expected occurrence, usually by some rational fraction of a beat) and/or ornamentation (inclusion of additional timepoints close to an expected timepoint).

\begin{figure}[h]
\centering
\includegraphics[width=.45\textwidth]{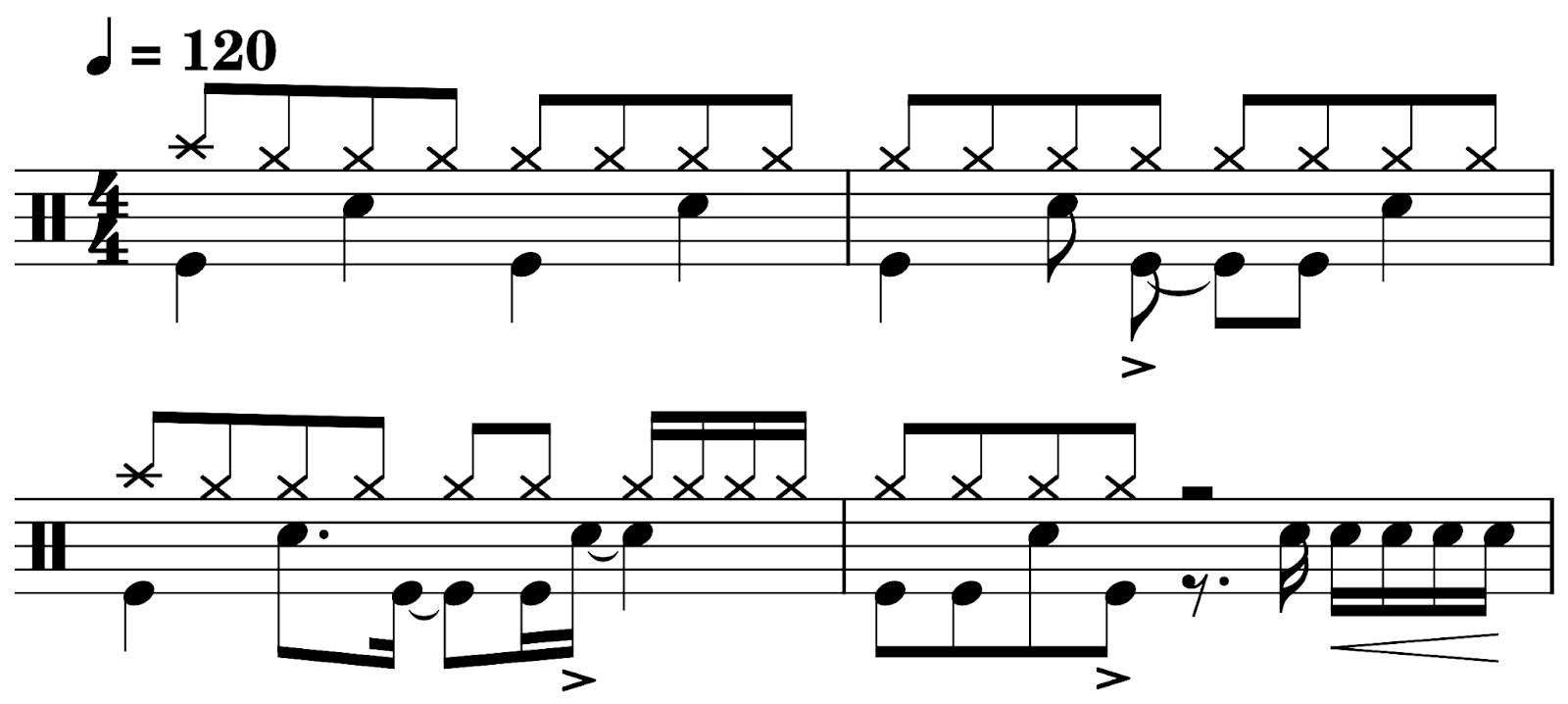}
\caption{Drumkit rhythm with increasing syncopations and ornaments}
\label{fig02}
\end{figure}

Consider the following example \ref{fig02} for basic drumkit---kick, snare, hihat, and crash. The first measure establishes the beat with kick and snare, while also establishing two-beat groupings with the kick drum, and an underlying eighth-note pulse with the hihat. In the second measure, that basic rhythm is elaborated with a single syncopation (the second kick note comes $\frac{1}{2}$ beat early) and a single ornamentation (insertion of another kick on the second half of beat 3). Those simple elaborations in the second measure are enough to give the rhythm a distinctive and more memorable character, while still retaining a close motivic relationship with the fundamental pattern of the first measure.

In the third measure, the amount of syncopation and ornamentation is increased slightly. Both the second kick note and the second snare note are now syncopated $\frac{1}{4}$ beat early (while retaining the kick ornament on the second half of beat 3, established in the previous measure), and the hihat is ornamented with inserted sixteenth notes in beat 4. It's noteworthy that the introduction of sixteenth-note divisions in this measure increases the underlying pulse, adding additional energy to the rhythm, and the recurrence of the crash on the downbeat suggests a two-measure (8-beat) grouping. The fourth measure shows what could be thought of as an extreme case of syncopation, a brief suppression of all expected events on beat 3. This is known as a ``stop'' in popular music, a well known technique for establishing anticipation by completely disappointing the listener's expectations for a short period, yet not for so long as to cause the listener to abandon the sense of beat and attendant expectations. The stop sets up in the listener an even stronger anticipation and desire for fulfillment on the next downbeat, which is reinforced by the sixteenth-note ``fill'' in the snare on beat 4.

\begin{figure}[h]
\centering
\includegraphics[width=.48\textwidth]{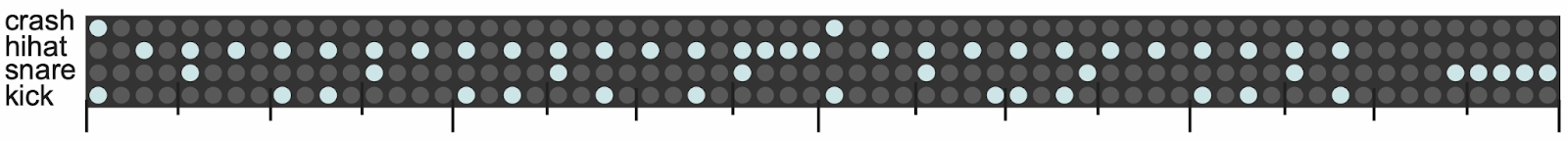}
\caption{Drumkit rhythm depicted on a temporal grid}
\label{fig03}
\end{figure}

If we think of these events as timepoints of occurrence on a temporal grid---a time ruler, if you will---of measures, beats, and sixteenth-note pulses, we can depict the entire musical passage in simplified form as just events and non-events, i.e., binary 1s and 0s \ref{fig03}. This can serve as a visual aid to perceiving periodicities, patterns, variations, changes in density, and the role of each distinct element (each instrument) in the overall contrapuntal texture. It also suggests that the compositional concepts of syncopation and ornamentation can be rethought and restated as deletion and insertion of events in a previously established---and perhaps strongly stylistic or motivic---pattern. As long as enough of the crucial skeletal timepoints are retained to remind the listener of the new rhythm's relationship to the previously heard pattern, a great many deletions and insertions can be made, with almost arbitrary variety, resulting in a great many elaborations of a rhythmic motive.

\begin{figure}[h]
\centering
\includegraphics[width=.25\textwidth]{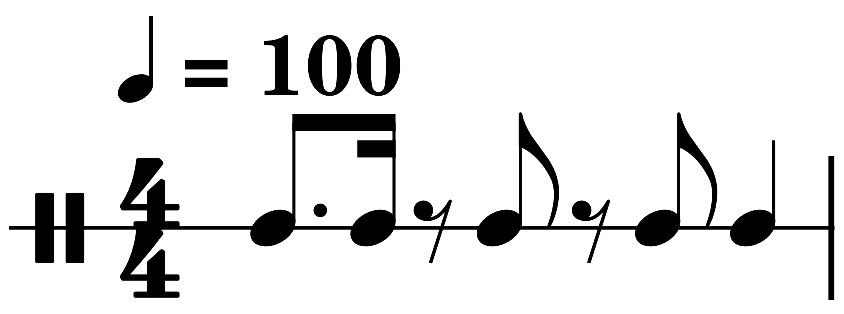}
\caption{The \textit{son clave}}
\label{fig04}
\end{figure}

For an example of quantum rhythm generation here, we will use that notion of rhythm as event vs. non-event (1 or 0) on a temporal grid. For our underlying recognizable motivic rhythm, we'll choose the \textit{son clave}, a well known rhythm in Afro-Cuban music \ref{fig04}. In temporal grid terminology, with a sixteenth-note grid resolution, this will yield the sixteen-bit pattern 1001001000101000.

\begin{figure}[h]
\centering
\includegraphics[width=.45\textwidth]{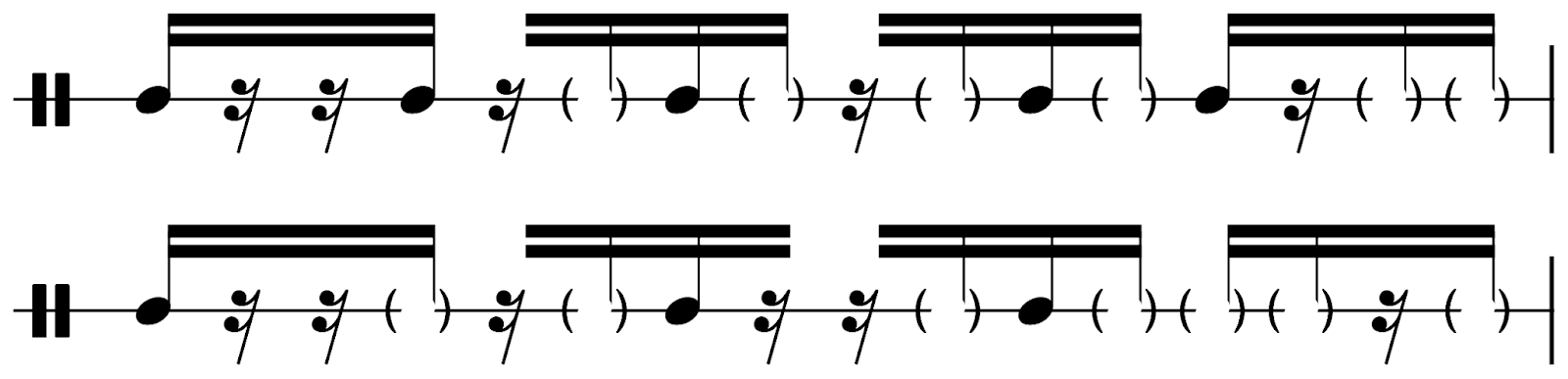}
\caption{Two measures of son clave, with some timepoints changeable}
\label{fig05}
\end{figure}

To enact the deletion and insertion operations, we'll leave certain timepoints in that pattern immutable, while designating others as potentially changeable. For every two-measure period, we'll leave 19 of the timepoints unchanged, and for the other 13 timepoints the quantum circuit will measure qubits to obtain either a 1 or a 0. We selected the changeable bits in such a way that most of the events in the clave are left intact, with only two of them being potentially deleted in the second measure \ref{fig05}. Thus, the 32-bit pattern for two measures of son clave, with changeable bits marked as X (and with spaces, to show each beat) is 1001 0x1x 0x1x 10xx $|$ 100x 0x10 0x1x xx0x.

With this method of rhythm generation, the algorithm will compose a continually changing rhythm, made up of over 8,000 different possible 8-beat patterns, all of which strongly evoke the fundamental son clave motive. 

\section{Timbre}\label{sec:timbre}
Whereas compositional decisions about rhythm take place at the note-level or phrase-level time scale, for digital sound synthesis we generally need to operate at the sub-millisecond level of individual audio samples (e.g., 48,000 samples per second). This comparatively large amount of data might be likened to the many measurements needed to track a quantum particle or to measure a quantum state.

\begin{figure}[h]
\centering
\includegraphics[width=.25\textwidth]{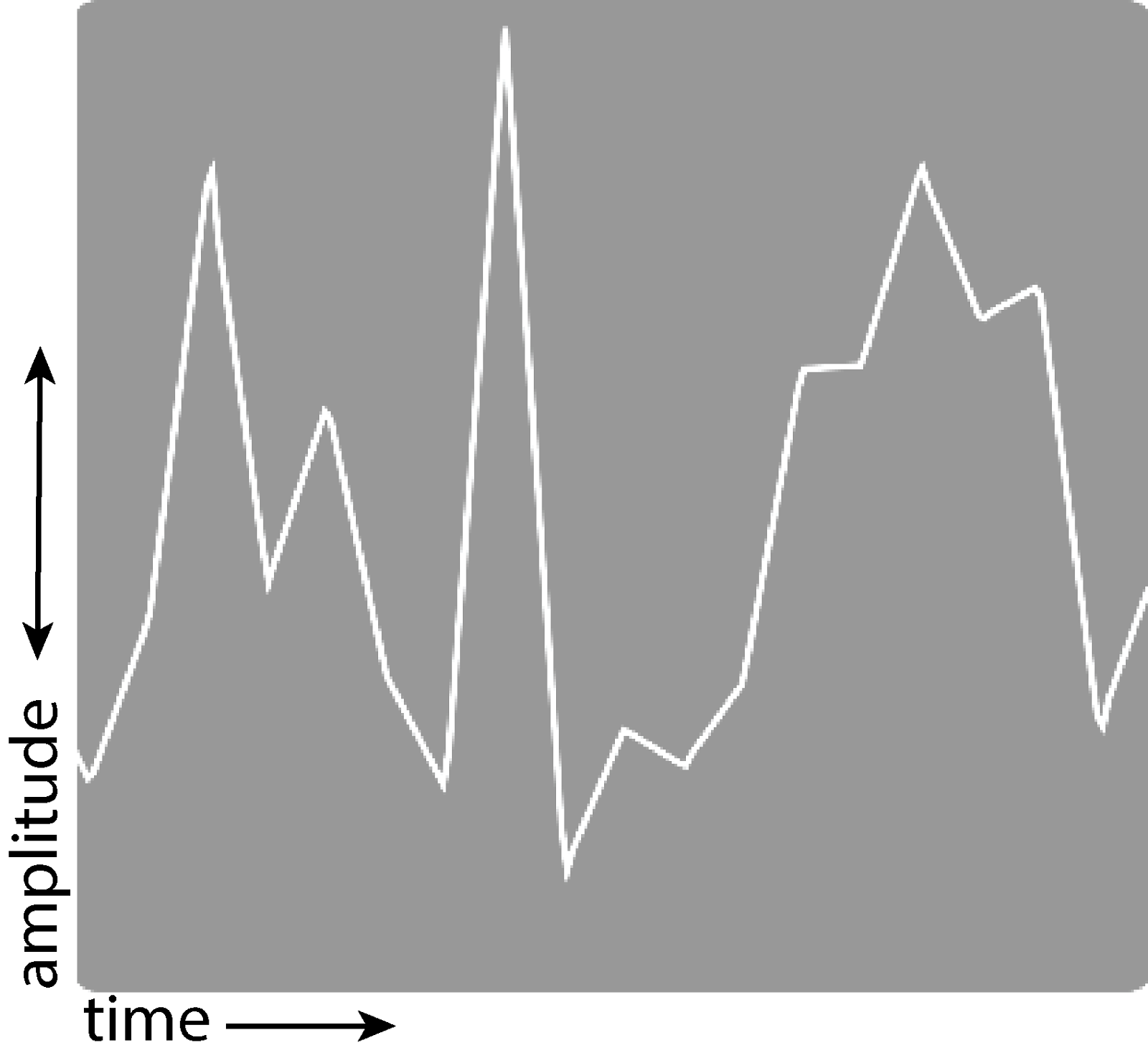}
\caption{Tracking one dimension of randomly changing motion, analogous to lowpass-filtered noise}
\label{fig06}
\end{figure}

When we think of a quantum particle's changing position in multi-dimensional space, we can presume that it is constantly moving, colliding with other particles, and changing direction. If we were to view it in such a way that we only consider its movement in one dimension, and we track that movement over time, we can plot its movement as a two-dimensional graph, with time on the $x$ axis and its displacement in that dimension on the $y$ axis. Its velocity will be represented by the slope of its displacement in our graph, and every change of direction will imply a collision. The result will be a graph of line segments from one random point to another. If we then slow that way down to such a rate that there are, say, 20 to 200 ``collisions'' per second \ref{fig06}, the sonic result will be very low-frequency filtered noise.

\begin{figure}[h]
\centering
\includegraphics[width=.25\textwidth]{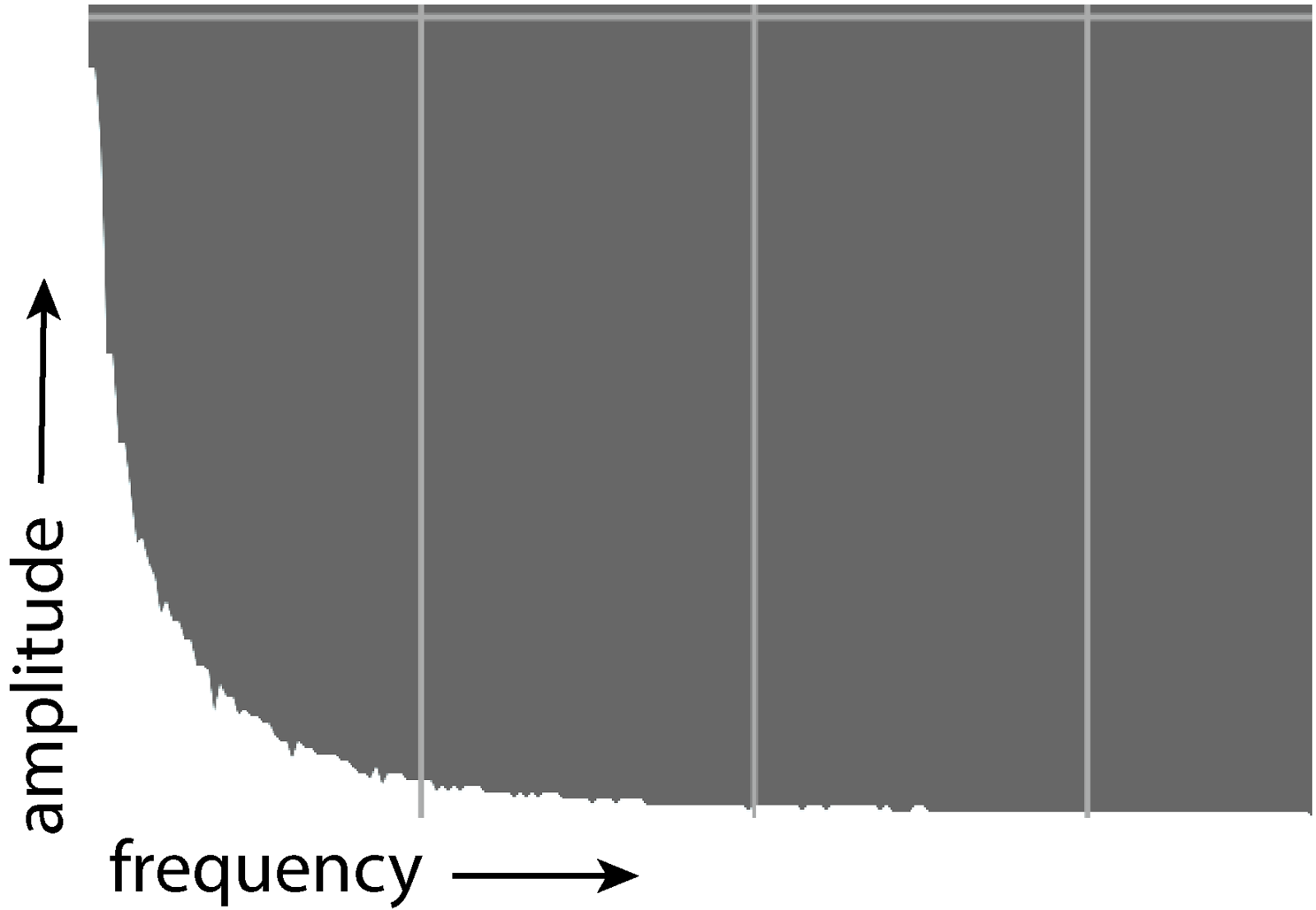}
\caption{Spectrum of lowpass-filtered noise}
\label{fig07}
\end{figure}

The spectrum of that noise is symmetrical, in both positive and negative frequencies, centered around 0 Hz \ref{fig07}.

\begin{figure}[h]
\centering
\includegraphics[width=.25\textwidth]{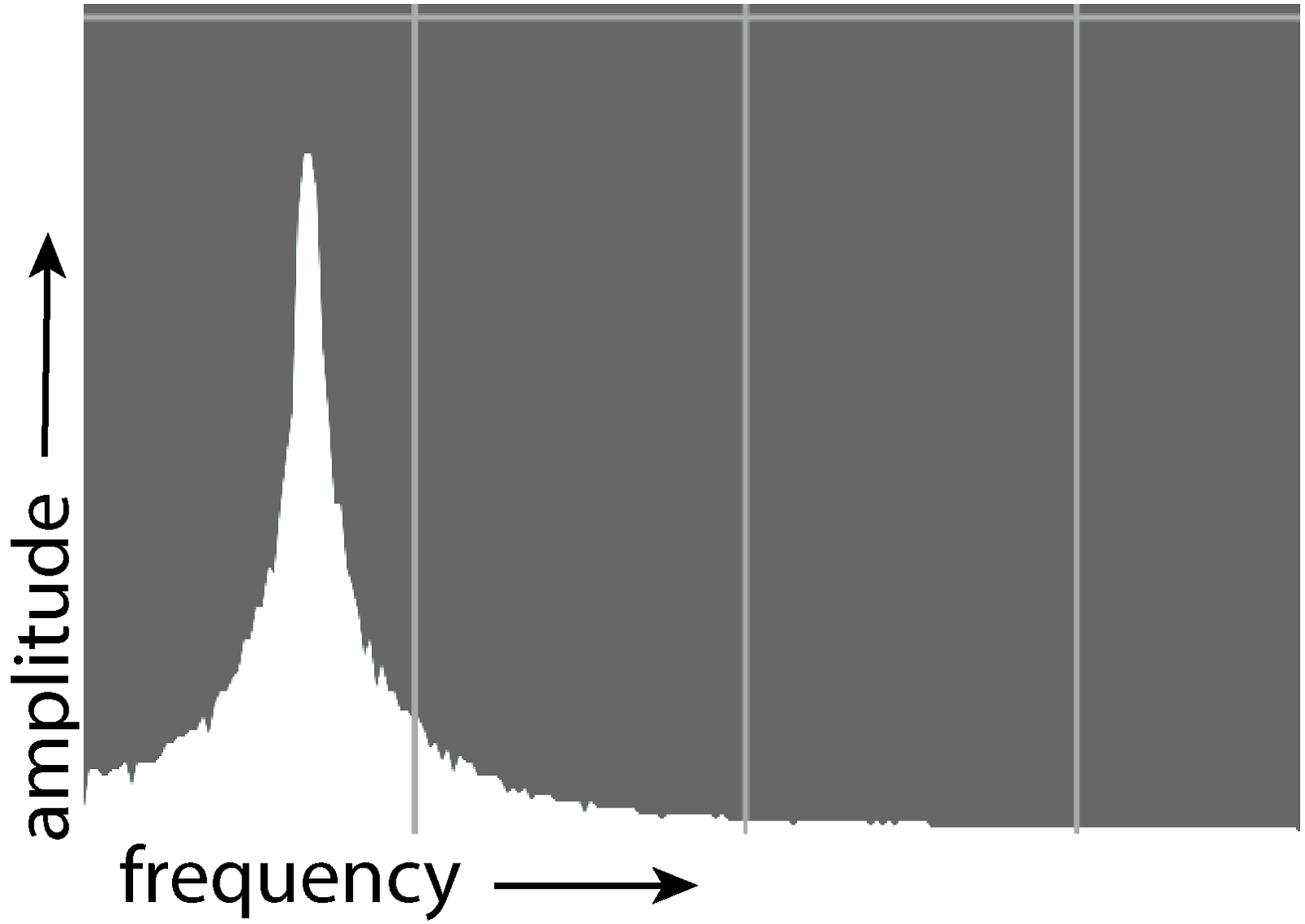}
\caption{Spectrum of lowpass-filtered noise, frequency-shifted}
\label{fig08}
\end{figure}

If we multiply that noise signal by a sinusoidal signal at an audible frequency---a process known in electronic music as \textit{ring modulation}---we obtain bandpass-filtered noise at that frequency \ref{fig08}, the bandwidth of which is determined by the average rate of change of direction in the random motion.

The resultant sound is a rather pleasing whistling effect, which can range from a very focused pitched tone to a noise band of barely determinable pitch. This sound can be used as the material for synthesized musical notes, which can sound long and sustained like a wind instrument or which can be made to sound percussive, like a xylophone or small drum, by windowing only brief instants of sound. We have implemented this idea as a versatile polyphonic synthesizer in the Max programming environment, using the QAC Toolkit \cite{hamidoQAC2022} to emulate the quantum circuit measurements. This is one approach to imagining ``the sound of quantum tracking'' as raw material for sound synthesis.

\section{Harmony}\label{sec:harmony}
In a previous article \cite{dobrianIntroQuantumHarmony2024}, we described the use of a quantum circuit emulation to make compositional decisions regarding the progression of chords---notably the diminished seventh chord, because of its ambiguity and its susceptibility to multiple interpretations and implications. We demonstrated that those chords, and others like them, function very much as if in a state of superposition, in the sense that their true musical meaning becomes evident to the listener only upon their resolution (i.e., their ``measurement''). In that article we showed musical examples produced by employing quantum circuits for chord choices, and another algorithm for voice leading, producing textbook-like four-voice homophonic choral harmonies. However, there are many other ways composers employ harmonic thinking---through counterpoint, arpeggiation, etc. We're currently exploring more quantum-inspired compositional techniques to generate new manifestations of harmonic progression, particularly by synthesizing stochastic polyphonic and granular textures.

In the mid-1950s, the composer Iannis Xenakis, observing the density and complexity of the highly contrapuntal atonal musical textures that were prevalent in western European avant garde classical music of the time, commented that, ``The enormous complexity prevents one from following the tangled lines and ... what will count will be the statistical average of isolated states of the components' transformations at any given moment.'' \cite{xenakisCriseMusiqueSerielle1955} This view led him to develop his own approach to free stochastic music, in which the attributes of individual notes were calculated as statistical distributions in a large number of events. In a comparable way, quantum computing uses large numbers of measurements of a quantum system to ``collapse'' its state into something classically knowable and useful.

\begin{figure}[h]
\centering
\includegraphics[width=.45\textwidth]{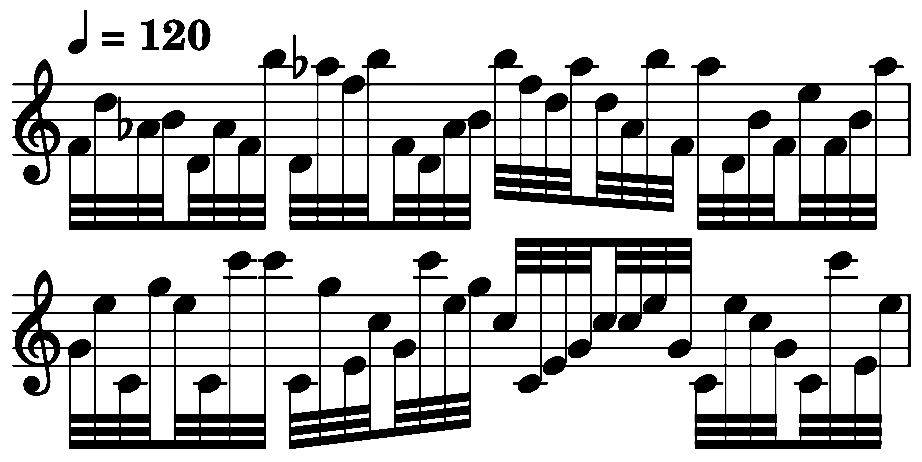}
\caption{Clouds of rapid notes with clear harmonic implications}
\label{fig09}
\end{figure}

As a way of reifying this conceptual parallel between stochastic music and quantum measurement, we designed an algorithm that composes musical textures in which the harmonic belongingness of each note to one chord or another is modeled as a single measurement of a quantum circuit. Because many measurements of the circuit are required to get a statistically accurate sense of the circuit's behavior, we use ``clouds'' of many musical notes to make a sonic texture with the desired harmonic implications. For example, a cloud of 32 notes, all of which are B, D, F, or Ab will very clearly imply a B diminished seventh chords, whereas following that with 32 more notes, all of which are C, E, or G will imply a resolution of the B diminished seventh chord to a C major triad \ref{fig09}. Performing those notes at a very rapid rate discourages listeners from making note-to-note melodic connections, and encourages cognition of the notes more as a harmonic texture; it also enables a larger sampling of notes to be presented in a shorter period of time.

\begin{figure}[h]
\centering
\includegraphics[width=.45\textwidth]{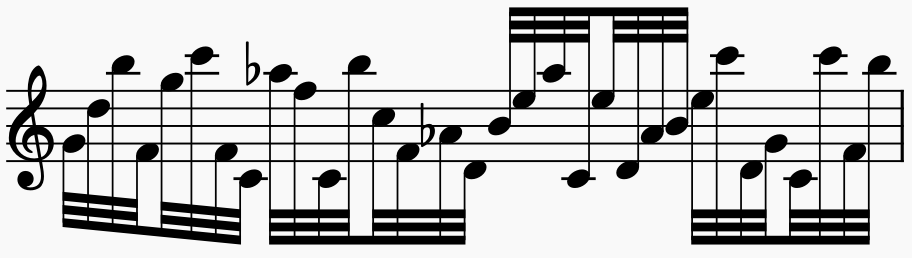}
\caption{Approximately equal probability of notes of Bdim7 or C chords}
\label{fig10}
\end{figure}

Using each measurement of a qubit, 1 or 0, to determine a note's membership in one of two chords, we obtain a cloud of notes that exemplifies (i.e., implies) those harmonies probabilistically by taking many successive measurements. A measurement of 0 results in a note of chord $A$, while a measurement of 1 results in a note of chord $B$. If the probability of 0 or 1 is approximately equal, the result will be a blend of those two harmonies \ref{fig10}. If those notes overlap due to sustain or reverberation (as if with the pedal held down on a piano), the harmonic blend becomes even more pronounced.

\begin{figure}[h]
\centering
\includegraphics[width=.45\textwidth]{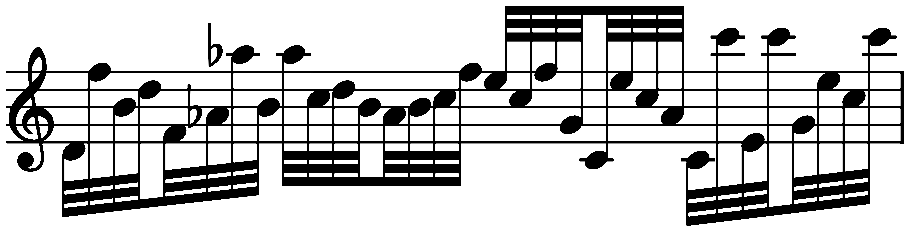}
\caption{Probabilistic crossfade during beats 2 and 3 from Bdim7 to C}
\label{fig11}
\end{figure}

If the probabilistic characteristic of the qubit's wave function changes over time, however, we hear a sort of ``harmonic crossfade'' from one predominant chord to another. For example, in a two-second passage of music, if there is a 100\% probability of chord $A$ for half a second (and thus a 0\% probability of chord $B$), then over the course of the next second those probabilities crossfade, until in the final half second there is a 100\% probability of chord $B$, the effect is a quick ``modulation'' from one harmonic predominance to another \ref{fig11}.

\begin{figure}[h]
\centering
\includegraphics[width=.25\textwidth]{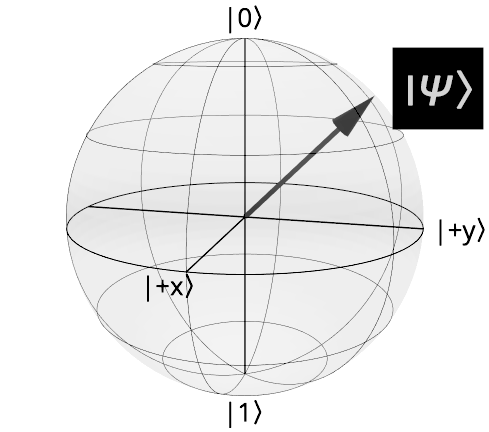}
\caption{Bloch sphere representation displaying an arbitrary state vector $\ket{\Psi}$}
\label{fig12}
\end{figure}

We know that in quantum computing the basis states of $\ket{0}$ and $\ket{1}$ can have unequal probabilities of being measured. Those probabilities can be transformed by rotating the state vector on the Bloch sphere representation of the qubit \ref{fig12} using an Rx gate. The rotation value for the Rx gate can be easily calculated for any desired probability $P$ using the equation:

\begin{equation}\label{eq1}
Rx = arccos(-2*P+1)
\end{equation}

Taking off from our previous work on computer-composed chord progressions using quantum circuits, we are now using the same sort of decision-making algorithm to choose implicit harmonies for clouds of notes, and we are using state vector rotation to vary statistical probabilities within those clouds and to effectuate harmonic crossfades in granular textures. These processes are automated to produce passages of music that migrate through different harmonies in a way that is musically coherent in the classical sense, yet where the implied harmonies are decided upon by a quantum-computing-aided technique.

\section{Spatialization}\label{sec:spatialization}

The perceived location of sounds in space has increasingly come to be considered an important parameter in recorded music and live computer music. Composers often consider the spatialization of sounds not merely as a special effect, but as an attribute of the music that can serve as an additional musical dimension and contribute as a meaningful component of the listener's aesthetic experience.

The fact that we have two ears enables us to pinpoint the directionality of sounds, primarily through interaural intensity difference (IID) and interaural time difference (ITD). Our sense of the sound's distance depends mainly on its amplitude relative to similar sounds we have heard previously, and also on the way that the sound reverberates in the space where it occurs. In this article we will be focusing on the compositional use of the sound's location, especially its direction as implied by IID.

Directionality in one dimension---e.g., left-to-right---can be readily achieved by the technique of intensity panning between stereo speakers. Simply put, the perceived azimuth angle of the sound is influenced by which of the two speakers is louder, because that discrepancy simulates the IID we experience in everyday acoustic sound perception. As a composer, one can control the perceived angle of the sound between the two speakers quite precisely by using an algorithm for constant-intensity panning \cite{mooreElementsComputerMusic1990, roadsComputerMusicTutorial2023}. Control of the sound's azimuth angle in two dimensions can be achieved by using four speakers placed as corners of a rectangular space, which permits panning both left-to-right and front-to-back. And for three-dimensional panning, the sound's perceived azimuth angle and elevation angle can be controlled by panning among eight speakers placed at the corners of a cube, permitting constant-intensity panning along $x$, $y$, and $z$ coordinates in Euclidean 3D space.\footnote{This is admittedly a vast oversimplification of the complex topic of surround sound. A great many other algorithms and speaker configurations are utilized for different applications. For the purpose of this discussion, however, we're deliberately restricting ourselves to IID constant-intensity left-to-right panning in stereo and to 3D constant-intensity panning using $xyz$ coordinates.}

Continuously changing the panning angle(s) of digital sound at the sample-by-sample level gives the illusion of the sound source moving continuously in space. The path the sound follows in virtual space is referred to as a \textit{soundpath}. Simple soundpaths such as a directional panning from side to side in stereo, or a circular panning in 2D space, are quite easily recognized by most listeners and thus can even be used as musical motives by a composer. For more complex sound paths, it's debatable how good most listeners are at discerning and remembering a soundpath well enough to compare one soundpath to another, because that is not a skill that most people develop. Nevertheless, even if one doesn't consciously hear soundpaths as musically meaningful, variations in location and movement can assuredly have an enlivening and gestural effect for listeners. Effects both subtle and fantastical have been used in popular music production as well as in experimental computer music.\footnote{It's noteworthy that high-velocity sound movement would realistically cause noticeable Doppler shift in the sound's perceived pitch, which might or might not be considered desirable, depending on the content of the music. Doppler shift is calculated as a changing delay time based on the virtual location's changing distance from the listener. We will not include that in this discussion.}

Another interesting approach to movement in spatialization is to place every sound event at a specific stationary location in virtual space, and to let changes in position from one event to the next create the effect of a soundpath. This approach has a few advantages. For each event, the relative gain of each speaker needs to be calculated only once. Stationary sound sources seem more ``realistic'', in the sense that most real-world acoustic music does not involve sound sources flying through the air; stationary sound locations give a sense of polyphony (multiplicity of sources) more than of movement. Using multiple stationary sources in lieu of moving sources also enables sound to leap discontinuously in space from one event to another, enabling different sorts of musical effects such as antiphony, spatial hocket, and the like. For the sorts of short percussive rhythmic articulations and many-note granular harmonic textures that we've discussed above, those polyphonic and antiphonal effects can add energy to the musical experience. The disadvantage of using many stationary virtual locations is that, if we want sound events in different locations to overlap in time, we need each sound to have its own unique gain adjustments before it is sent to the loudspeakers; therefore, the software generating the sounds must be truly polyphonic in its panning capabilities.

To employ a quantum computing phenomenon in the spatialization of sound, we wrote an algorithm that emulates quantum particle tracking ``measurement error'' and uses those deviations to alter the virtual location of sound sources. In this way, the measurement error adds a subtle dislocation of each sound from a single ostensible location, or can be used to perturb an otherwise stable soundpath, creating a miniature random subpanning within a larger soundpath.

\begin{figure}[h]
\centering
\includegraphics[width=.25\textwidth]{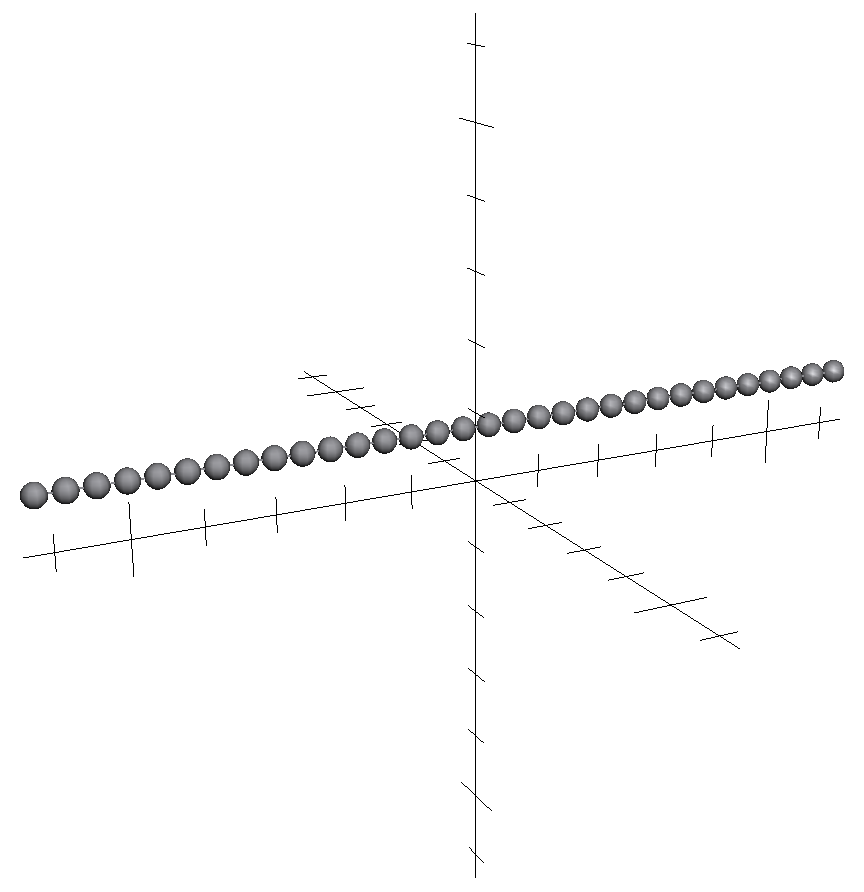}
\caption{Linear soundpath implied by a series of stationary sound events}
\label{fig13}
\end{figure}

To demonstrate this idea, let's consider a series of sound events, each of which is itself stationary, but which, in their aggregate, outline a linear or curvilinear soundpath. For simplicity, we'll choose thirty-two event locations that describe a constant-velocity linear progression from left to right \ref{fig13}.

\begin{figure}[h]
\centering
\includegraphics[width=.25\textwidth]{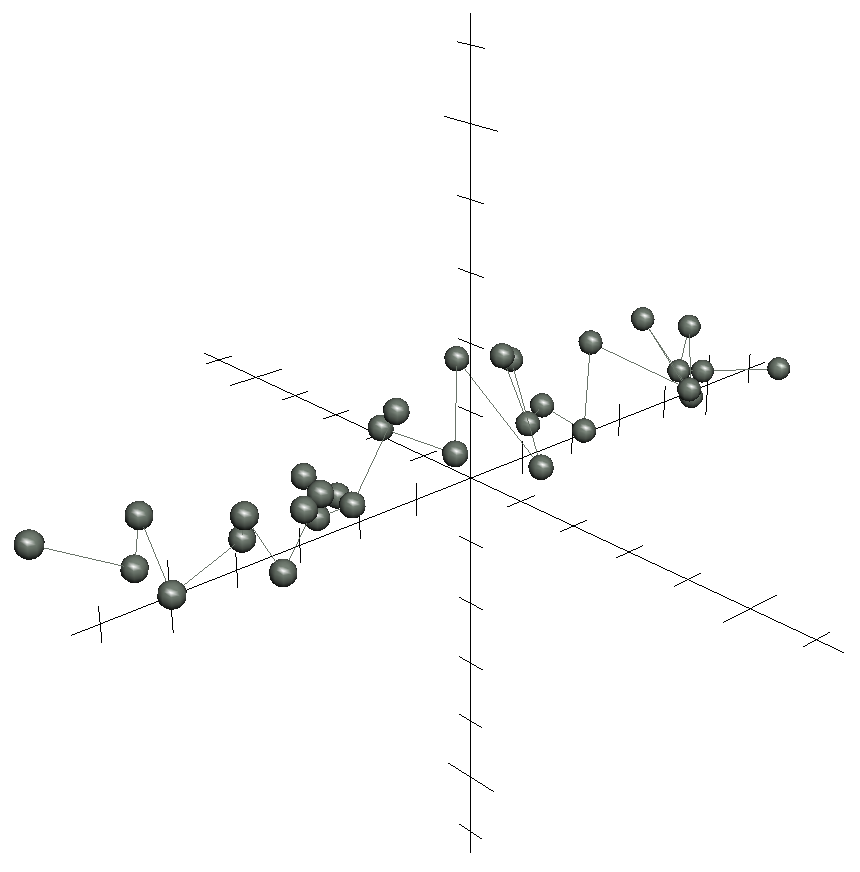}
\caption{Perturbed linear progression of event locations}
\label{fig14}
\end{figure}

If we take into account the ``measurement error'' that perturbs our perception of that implied soundpath, the actual panning of the sound events will be less predictable and more varied, and thus potentially a bit more complex and interesting aesthetically, yet will nevertheless strongly evince the original soundpath \ref{fig14}.

\begin{figure}[h]
\centering
\includegraphics[width=.25\textwidth]{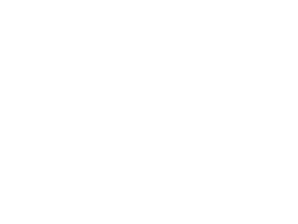}
\caption{Increasing and decreasing dispersity of event locations}
\label{fig15}
\end{figure}

Variations in the degree of perturbation can produce more or less dispersed spatial distribution of events \ref{fig15}.

All of the aforementioned compositional techniques can be successfully combined to create music with comprehensible and pleasing changes. Notes synthesized from ``quantum particle tracking'' filtered noise, with pitches that produce a probabilistic harmonic crossfade from a chord in a state of ``superposition'' to a chord of more recognizable identity, may be played in ornamented and syncopated rhythmic patterns, enhanced by algorithmically-derived patterns of panning. These techniques, individually and in combination, yield a significant variety of algorithmic music possibilities, each inspired by and implemented with ideas from quantum computing.

\section{Conclusion}
This is an abbreviated version of the paper that we intend to publish and present at ISQCMC 2025. It describes some promising explorations into the use of quantum computing to develop new techniques of computer-aided composition, which in turn potentially lead to new paradigms for thinking of the compositional process and thus to new musical experiences. Just as quantum computing has allowed us to rethink computation from its very basic foundations, we are motivated to reinvestigate features of music composition in areas such as rhythmic syncopation and ornamentation, the synthesis of musical timbres, the composition and realization of harmonic progressions, and the subtle variation of sound localization, and in so doing, rethink each of those features in the light of a new quantum-computing-aided composition paradigm.

{\centering All software examples and sound examples referred to in this article can be found at:
\href{https://harmony.quantumland.art/isqcmc25}{https://harmony.quantumland.art/isqcmc25} \par}

\noindent\textbf{Acknowledgments.} \anonymize{The work by Omar Costa Hamido is supported by the European Project
IIMPAQCT, under grant agreement Nr. 101109258.}

\begin{figure}[h]
\centering
\includegraphics[width=.25\textwidth]{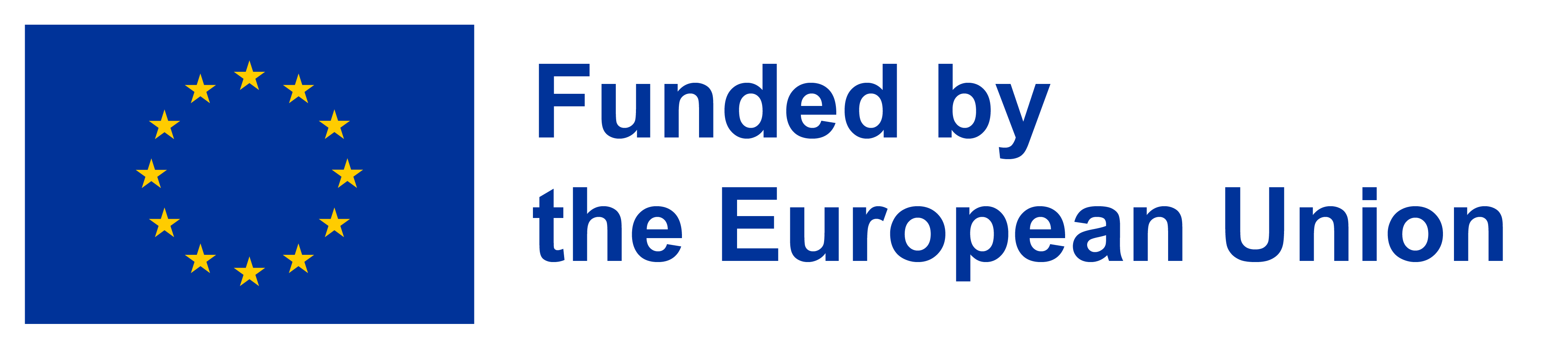}
\end{figure}

\bibliographystyle{unsrt}
\bibliography{references}

\end{document}